\documentclass[conference]{IEEEtran}
\IEEEoverridecommandlockouts
\usepackage{cite}
\usepackage{amsmath,amssymb,amsfonts}
\usepackage{algorithmic}
\usepackage{graphicx}
\usepackage{url}
\usepackage{hyperref}
\usepackage{textcomp}
\usepackage{array}
\usepackage{xcolor}
\newcommand{\aap}{Astronomy \& Astrophysics}
\def\BibTeX{{\rm B\kern-.05em{\sc i\kern-.025em b}\kern-.08em
    T\kern-.1667em\lower.7ex\hbox{E}\kern-.125emX}}
\begin{document}

\title{Simulating Ionospheric Distortions in Mock SKA-Low Solar Radio Observations \\

}

\author{
  \IEEEauthorblockN{1\textsuperscript{st} Rohit Sharma}
  \IEEEauthorblockA{\textit{Department of Space, Planetary \& Astronomical} \\ 
  \textit{Sciences \& Engineering (SPASE)} \\
  \textit{Indian Institute of Technology Kanpur}\\
 \textit{ Kalyanpur, Kanpur, 208016, U.P., India} \\
  rsharma@iitk.ac.in \\ ORCID: \url{https://orcid.org/0000-0003-0485-7098}}
  \and
    \IEEEauthorblockN{2\textsuperscript{nd} Mitin}
   \IEEEauthorblockA{\textit{Department of Space, Planetary \& Astronomical} \\ 
  \textit{Sciences \& Engineering (SPASE)} \\
  \textit{Indian Institute of Technology Kanpur}\\
 \textit{ Kalyanpur, Kanpur, 208016, U.P., India} \\
  mitin25@iitk.ac.in \\ ORCID: \url{https://orcid.org/0009-0006-9965-3176}
  }}

\maketitle

\begin{abstract}
The Square Kilometre Array Observatory (SKAO) will provide unprecedented angular resolution for low-frequency solar radio imaging, enabling access to fine-scale coronal structures. However, the improved spatial resolution of SKA-Low also increases the sensitivity of solar images to ionospheric propagation effects. We investigate the impact of ionospheric fluctuations on simulated SKA-Low solar observations at 50, 100, 150, and 200 MHz using the \textsc{Karabo} simulator and turbulent, moving ionospheric phase screens. We quantify the resulting image distortions using spatial power spectral density (PSD) analysis and compare the SKA response with corresponding MWA simulations. The results show that ionospheric distortions are strongly scale- and frequency-dependent, with the maximum fractional PSD distortion decreasing with observing frequencies. SKA resolves ionospheric fluctuations over spatial scales that are largely smoothed by the MWA instrumental response, with significant distortions occurring on scales comparable to solar active-region structures. These results demonstrate that ionospheric propagation effects should be considered when interpreting high-resolution SKA-low solar radio images, particularly at the lower end of the frequency band.

\end{abstract}


\section{Introduction}

The forthcoming Square Kilometre Array Observatory (SKAO) will provide unprecedented spatial, temporal, and spectral capabilities for solar radio observations across SKA-Low (50--350 MHz) and SKA-Mid (350 MHz--15 GHz) \cite{b5}. In particular, the long baselines and dense $uv$-coverage of SKA-Low are expected to improve the angular resolution of solar radio images by $\sim10$--$20$ times compared with precursor facilities such as the Murchison Widefield Array (MWA) \cite{b3,b4}, enabling access to fine-scale coronal structures that are currently unresolved. However, this improved resolution also makes SKA-Low increasingly sensitive to propagation effects. At low frequencies, radio waves from the Sun propagate through the turbulent, dispersive ionosphere, which introduces direction-dependent phase fluctuations and can produce apparent position shifts, brightness distortions, and scale-dependent spatial filtering. These effects may become particularly important for solar imaging because the intrinsic solar emission contains structures over a broad range of angular scales, from the full solar disk and active regions to fine coronal loops and compact radio sources.

In this work, we investigate the impact of ionospheric fluctuations on simulated SKA-Low solar observations and identify the spatial scales most affected by ionospheric propagation. We simulate solar radio observations using the full SKA-Low configuration at 50, 100, 150, and 200 MHz, with a turbulent, moving ionospheric phase screen incorporated into the simulations. By comparing images and spatial power spectra (PSD) with and without the ionosphere, we quantify the scale-dependent changes in the recovered solar brightness distribution. We further compare the SKA results with corresponding MWA simulations to examine how the different instrumental transfer functions and $uv$-sampling influence the visibility of ionospheric distortions. This analysis provides an estimate of the spatial scales and frequency range over which ionospheric effects may limit the interpretation of high-resolution SKA-Low solar radio observations.

%In the coming years, the under-construction Square Kilometre Array Observatory (SKAO) will offer unprecedented spatio-temporal-spectral data products in low (50 MHz-350 MHz) and high bands (350 MHz - 15 GHz) respectively \cite{b5}. Compared to its precursors like the Murchison Widefield Array (MWA), SKAO will significantly improve ($\sim$ 10-20 times) the spatial resolution of the solar radio images \cite{b3,b4}. However, the high-resolution solar radio images at low frequencies will be distorted by the ionosphere with complicated direction-dependent phase shifts. 

%In this paper, we study the impact of ionospheric distortion on simulated SKA solar observations. We study the spatial scales vulnerable to the ionospheric distortions at SKA-low frequencies for the full SKA-low configuration in the simulated solar radio disk maps.

\section{Simulation of Radio Solar Observations using Ionospheric Screens}

We simulate solar radio observations to investigate the impact of ionospheric fluctuations on the reconstructed solar images. The simulations are performed using the \textsc{Karabo} software \cite{b1}, which enables the generation of synthetic interferometric observations and images for different telescope configurations. We focus on the lower-frequency range of SKA-Low and consider four representative observing frequencies: 50, 100, 150, and 200 MHz. This frequency range allows us to investigate the frequency dependence of ionospheric distortions and their impact on the spatial structure and image fidelity of solar radio observations.

\subsection{Solar Model}
We adopt the thermal solar disk as the input sky model for the simulations, with the spatial distribution of the solar radio emission generated using the FORWARD model \cite{b2}. The model represents the solar radio emission under the assumption that thermal free--free (bremsstrahlung) processes dominate the emission at the frequencies considered here. The resulting solar brightness distribution is generated at a reference frequency of 240 MHz and subsequently used as the input sky model for the interferometric simulations.

To construct the sky models at other observing frequencies, we extrapolate the flux density from the 240 MHz reference model using a frequency-dependent scaling relation,
$S_{\nu}(\nu)
=
S_{\nu}(\nu_{\rm ref})
\left(\frac{\nu}{\nu_{\rm ref}}\right)^2,
$
where $\nu_{\rm ref}=240$ MHz and $S_{\nu}(\nu_{\rm ref})$ is the flux density of the reference solar model. Thus, the flux density at an observing frequency $\nu$ is given by
\begin{equation}
S_{\nu}
=
S_{240}
\left(\frac{\nu}{240~{\rm MHz}}\right)^2.
\end{equation}
This scaling corresponds to the optically thick Rayleigh--Jeans limit, and hence the flux density for a fixed angular source size scales as $S_{\nu}\propto\nu^2$. The same spatial morphology is therefore retained across the simulated frequency range, while the overall flux-density normalisation is adjusted according to the adopted frequency scaling. This provides a consistent set of solar input models for investigating the frequency-dependent effects of the instrumental response and ionospheric propagation.

\subsection{Ionospheric Model}

We simulate two vertically separated, moving ionospheric phase screens, each with a physical extent of $200\times200~\mathrm{km}$. The parameters adopted for the two ionospheric layers are summarised in Table~\ref{tab:ionospheric_angular_scales}. Each screen is sampled at a spatial resolution of $100~\mathrm{m,pixel^{-1}}$, and the two layers are separated by $10~\mathrm{km}$ in altitude. The combined screen extent corresponds to an angular field of view of approximately $40^\circ$ on the sky.

The simulation provides the line-of-sight variation in total electron content, $\Delta\mathrm{TEC}$, which is converted into an ionospheric phase perturbation according to
\begin{equation}
\phi_{\rm ion}
=
-\frac{8.44\times10^{9}\,\Delta\mathrm{TEC}}{\nu}.
\end{equation}
where $\Delta\mathrm{TEC}$ is expressed in TECU, with $1~\mathrm{TECU}=10^{16}~\mathrm{electrons,m^{-2}}$, and $\nu$ is the observing frequency in Hz. The negative sign represents the phase advance introduced by the ionospheric plasma.

The simulated turbulent structures span angular scales of approximately $0.1^\circ$, with a sampling interval of $\sim1'$ on the sky. These scales overlap with the characteristic angular scales of solar structures across the solar disk, allowing us to investigate the scale-dependent modification of solar radio images by ionospheric propagation effects.

\begin{table}[htbp]
\centering
\caption{Angular scales corresponding to the physical scales of the ionospheric screen, assuming a screen height of $h=300~\mathrm{km}$. Simulated using the ARAtmospy example set.}
\begin{tabular}{lcc}
\hline
\textbf{Quantity} & \textbf{Physical scale} &\textbf{$\theta$ / $k$ (arcmin$^{-1}$)} \\
\hline
Sampling & $0.1~\mathrm{km}$/pixel & $1.15'$/$5.46$ \\
Turbulence scale, $r_0$ & $5~\mathrm{km}$ & $57.3'/ 0.1$ \\
Max. screen scale, $b_{\max}$ & $20~\mathrm{km}$ & $3.82^\circ / 0.027$ \\
Screen width & $200~\mathrm{km}$ & $38.2^\circ / 0.0027$ \\
\hline
&\textbf{Layer 1}&\textbf{Layer 2} \\
\hline
Velocity & $150~\mathrm{km\,h^{-1}}$ & $75~\mathrm{km\,h^{-1}}$ \\
Direction & $60^\circ$  & $-30^\circ$ \\
Layer height & $300~\mathrm{km}$ & $310~\mathrm{km}$ \\
\hline
\textbf{SKA Frequency} & \textbf{$\Delta_{max}$} &\textbf{$\theta_{max}$ / $k_{max}$ (arcmin$^{-1}$)} \\
\hline
50 MHz & 0.45 & $3.14'$/$2.0$ \\
100 MHz & 0.39 & $1.79'/ 3.5$ \\
150 MHz & 0.31 & $1.23' / 5.1$ \\
200 MHz & 0.09 & $1.39' / 4.5$ \\
\hline
\textbf{Solar Scales} & \textbf{$\theta$}  &\textbf{$k_{max}$ (arcmin$^{-1}$)} \\
\hline
Solar disk & $\geq 30'$ & $\leq 0.21$ \\
Active Regions & $1.67'$ & $3.77$ \\
Fine loops & $0.17'$ & $37.70$ \\
\hline
\end{tabular}
\label{tab:ionospheric_angular_scales}
\end{table}

\begin{figure*}[h]
\centerline{\includegraphics[width=0.95\textwidth]{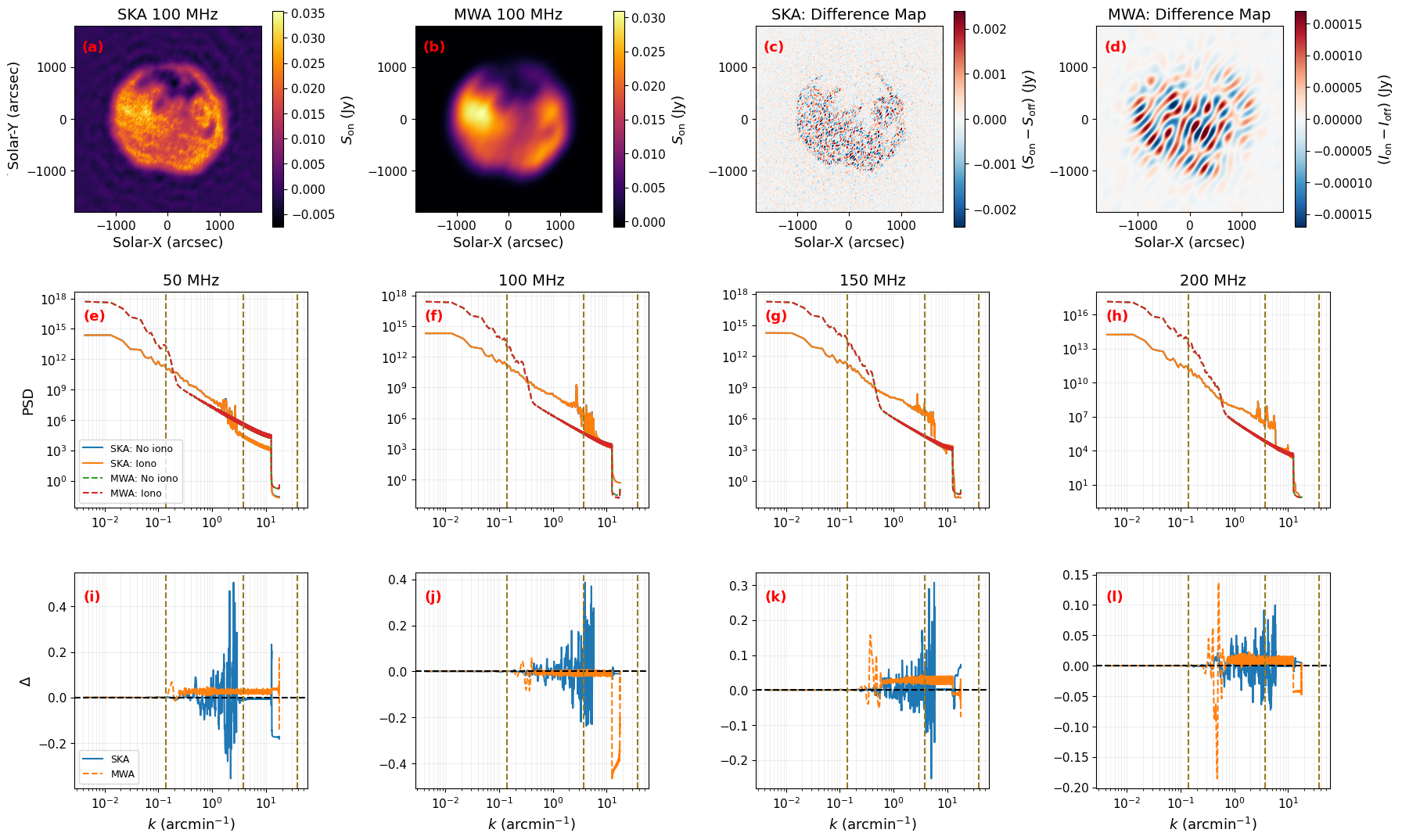}}
\caption{Simulated solar images and computed PSD of the maps with and without ionospheric screens. Panel (a) \& (b) show the solar maps at 100 MHz with an ionospheric screen from SKA and MWA configurations, respectively. Panels (c) and (d) show the difference in the flux densities with and without ionospheres. Panels (e), (f), (g) \& (h) show the PSD for SKA and MWA simulated maps with and without an ionosphere for selected four SKA-low frequencies. The last row quantifies the differences in the PSD due to switching on an ionosphere for SKA and MWA configurations. }
\label{fig}
\end{figure*}

\section{Power spectrum of the Ionospheric Fluctuations}

We quantify the spatial distribution of the solar radio emission in terms of the flux density, $S_{\nu}(x,y)$, where $(x,y)$ denotes the angular coordinates in the radio image. The two-dimensional Fourier transform of the flux-density distribution, $\widetilde{S}*{\nu}(k_x,k_y)$, is used to characterise the spatial power distribution through the two-dimensional power spectral density (PSD), $P(k)=|\widetilde{S}*{\nu}(k_x,k_y)|^2$, where $k^2=k_x^2+k_y^2$. Assuming approximate radial symmetry in the spatial power distribution, we azimuthally average the two-dimensional PSD to obtain the radial PSD, $P(k)$. The corresponding angular scale is related to the spatial frequency through $\theta \sim 2\pi/k$, such that larger $k$ values correspond to progressively finer angular structures.

The measured PSD of a radio-interferometric image contains contributions from the intrinsic spatial structure of the solar emission as well as the instrumental response of the interferometer. In the presence of propagation effects, the observed spatial power can be approximately expressed as
\begin{equation}
P(k,\nu) \approx P_{\odot}(k,\nu)
\left|H_{\rm inst}(k,\nu)\right|^2
\left|H_{\rm ion}(k,\nu)\right|^2,
\end{equation}
where $P_{\odot}(k,\nu)$ represents the intrinsic solar brightness distribution, $H_{\rm inst}(k,\nu)$ is the instrumental transfer function determined by the interferometer response and $uv$-sampling, and $H_{\rm ion}(k,\nu)$ represents the effective transfer function associated with the ionospheric phase screen. This formulation provides a convenient framework for separating the effects of the intrinsic solar morphology, instrumental sampling, and ionospheric propagation in the simulated observations.

For the two interferometric configurations considered here, the corresponding PSDs can be written as
\begin{equation}
P_{\rm SKA}(k,\nu) =
P_{\odot}(k,\nu)
\left|H_{\rm SKA}(k,\nu)\right|^2
\left|H_{\rm ion}(k,\nu)\right|^2,
\end{equation}
and
\begin{equation}
P_{\rm MWA}(k,\nu) =
P_{\odot}(k,\nu)
\left|H_{\rm MWA}(k,\nu)\right|^2
\left|H_{\rm ion}(k,\nu)\right|^2.
\end{equation}
Because both the instrumental response and the ionospheric propagation effects are frequency dependent, we retain the explicit frequency dependence in the PSD and transfer functions.

To quantify the change in spatial power introduced by the ionosphere, we define the fractional PSD residual as
\begin{equation}
\Delta^{\rm ion}_{\rm inst}(k,\nu)
=
\frac{
P^{\rm ion}_{\rm inst}(k,\nu)
-
P^{\rm no-ion}_{\rm inst}(k,\nu)
}{
P^{\rm no-ion}_{\rm inst}(k,\nu)
}.
\end{equation}
where ${\rm inst}$ denotes either SKA or MWA. This quantity measures the fractional change in spatial power produced by introducing the ionospheric phase screen while keeping the underlying solar emission and instrumental configuration unchanged. Thus, $\Delta^{\rm ion}*{\rm inst}>0$ indicates an enhancement of spatial power at the angular scale $\theta\sim2\pi/k$, whereas $\Delta^{\rm ion}*{\rm inst}<0$ indicates a suppression of power at that scale. The sign of $\Delta^{\rm ion}_{\rm inst}$ therefore distinguishes between scale-dependent power enhancement and suppression caused by the redistribution of solar brightness by ionospheric propagation. These changes reflect the combined effects of ionospheric phase fluctuations, scattering, and spatial filtering, with the observed response further modulated by the baseline distribution and $uv$-sampling of the interferometer.

Figure~\ref{fig} summarises the resulting ionospheric distortions and their corresponding spatial PSDs at 50, 100, 150, and 200 MHz for the SKA and MWA configurations. The figure demonstrates how the ionospheric response depends on observing frequency, spatial scale, solar morphology, and interferometric $uv$-sampling. In particular, comparison of the ionosphere-on and ionosphere-off PSDs allows us to identify the spatial scales over which the ionosphere modifies the recovered solar structures for each instrument.

Table~\ref{tab:ionospheric_angular_scales} summarises the characteristic ionospheric impact by listing the value of $\Delta^{\rm ion}$ and the corresponding spatial frequency at which the maximum PSD distortion occurs. We denote these quantities by $\Delta_{\max}$ and $k_{\max}$, respectively, as identified from the second row of Fig.~\ref{fig}. These parameters provide a compact measure of both the amplitude and characteristic spatial scale of the ionospheric distortion at each observing frequency and for each interferometric configuration.

\subsection{Morphological Variations}

Fig.~\ref{fig} (panels a--d) shows the spatial distribution of the solar-disk brightness at 100 MHz, together with the corresponding brightness distortions introduced by the ionosphere. Owing to its higher angular resolution, SKA produces a substantially sharper solar disk than MWA. In the SKA residual image (panel c), the ionospheric contribution produces brightness distortions of approximately $1$--$10\%$. These residuals are spatially well localised, with the distortions closely associated with the underlying structures in the solar map (panel a), demonstrating the ability of SKA to resolve the spatial structure of the ionospheric perturbations. In contrast, the MWA residual map exhibits weaker but more extended, large-scale variations, indicating that its instrumental response smooths the smaller-scale ionospheric fluctuations. Consequently, the ionospheric distortions are more readily resolved in the SKA images.

The corresponding SKA PSD residuals (panel i) show enhanced power over spatial frequencies $0.1 < k < 5$ arcmin$^{-1}$, broadly overlapping with the turbulent spatial scales introduced in the simulated ionospheric screen. The observed PSD, however, contains contributions from both the ionospheric fluctuations and the intrinsic spatial structure of the solar brightness distribution. In the second and third rows, the three vertical dashed lines indicate characteristic solar spatial scales: the solar-disk scale ($0.5^\circ$, $k=0.2$ arcmin$^{-1}$), active-region scale ($100''$, $k\approx4$ arcmin$^{-1}$), and fine loop/source scale ($10''$, $k\approx37$ arcmin$^{-1}$). We find that, at 100 MHz, ionospheric fluctuations primarily affect spatial scales extending from the solar-disk scale to scales comparable to or smaller than active regions in the SKA images. This impact is inherently frequency dependent, with the characteristic affected spatial scales and distortion amplitudes changing with observing frequency.

\subsection{Frequency dependence on the fluctuations}

Within the SKA-Low frequency range, ionospheric distortions are expected to decrease approximately as $1/\nu^2$, while the characteristic spatial scales affected by the ionosphere shift toward higher spatial frequencies ($k$) with increasing observing frequency. This trend is evident in Fig. \ref{fig} (panels e--h), where the ionospherically affected fluctuations in the SKA PSD progressively shift toward smaller spatial scales at higher frequencies. As shown in Table~\ref{tab:ionospheric_angular_scales}, the brightness distortion, $\Delta$, decreases substantially from $\sim45\%$ at 50 MHz to $\sim9\%$ at 200 MHz. The significantly reduced distortion at 200 MHz indicates that the ionospheric impact becomes comparatively weak toward the upper end of the SKA-Low frequency band. Overall, this frequency dependence is consistent with the expected wavelength dependence of ionospheric refraction and scattering, which become progressively weaker at higher observing frequencies.

\section{Discussion and Summary}

SKA's higher angular resolution and extended baselines provide improved image dynamic range and fidelity compared with precursor instruments such as MWA. The ionospheric impact is weak and broadly comparable between SKA and MWA at large spatial scales, while a modest differential response becomes apparent at intermediate spatial scales. At smaller spatial scales, SKA is able to resolve structures on which ionospheric distortions become increasingly visible, whereas the MWA instrumental transfer function suppresses or smooths these spatial modes. By comparing the PSDs with the ionosphere switched on and off, we qualitatively investigate the corresponding transfer functions, $H_{\rm ion}(k,\nu)$ and $H_{\rm SKA}(k,\nu)$. The maximum spatial frequencies accessible to SKA, $k_{\max}$, extend beyond the characteristic scales of ionospheric turbulence ($k>0.1$ arcmin$^{-1}$) and overlap with the spatial scales of solar active regions (Table~\ref{tab:ionospheric_angular_scales}). At 50 MHz, even under the strongest ionospheric distortions considered here, SKA can recover solar structures down to $\approx17$ arcsec \cite{b6}; however, the ionosphere introduces up to $\sim45\%$ power distortion at spatial scales below $\sim3$ arcmin. In realistic 50 MHz solar observations, coronal scattering is expected to further broaden the apparent sizes of solar structures, particularly on scales $\gtrsim30''$. Consequently, the effects of ionospheric distortion may be shifted toward larger angular scales in the observed images, making the combined impact of ionospheric and coronal propagation effects important when interpreting fine-scale solar structures. The combined effects of system noise and other instrumental responses on the reconstructed solar images remain to be quantified and will be addressed in future work.

\section*{Acknowledgment}
RS acknowledge Swiss-SRC colleagues, especially the Karabo team, for fruitful discussions. RS also acknowledge IIT Kanpur for continuing support for SKA-related activities.

\end{document}